%% file: main.tex
\documentclass[10pt,conference, letterpaper]{IEEEtran}
\IEEEoverridecommandlockouts

\usepackage{configuration}

\begin{document}
\bstctlcite{IEEEexample:BSTcontrol}

\title{Adaptive RIS-aided Communications through 
ML-based Generation of Phase Masks}
\author{
  \IEEEauthorblockN{
    Corwin Carpenter\IEEEauthorrefmark{1},
    Thomas Daltzis\IEEEauthorrefmark{2}
    George C. Trichopoulos\IEEEauthorrefmark{2}
    Jacek Kibilda\IEEEauthorrefmark{1}, and 
    Joao F. Santos\IEEEauthorrefmark{1}, 
  }
  \IEEEauthorblockA{
    \IEEEauthorrefmark{1}\textit{Commonwealth Cyber Initiative},
    \textit{Virginia Tech}, USA,
    e-mail: \{ccarpenter, jkibilda, joaosantos\}@vt.edu}
    \IEEEauthorrefmark{2}\textit{Arizona State University}, USA,
    e-mails: \{tdaltzis, gtrichop\}@asu.edu
}
\maketitle

\begin{abstract}
\acp{RIS} are an attractive technology for \ac{mmWave} communications due to their ability to passively reflect incident signals.
However, current implementations of \ac{RIS} rely on performing computationally-intensive algorithms offline to generate phase masks, which are stored as a codebook on the embedded microcontroller on the \ac{RIS}. The codebook size is restricted by the embedded microcontroller's storage capacity, which limits the ability of the \ac{RIS} to adapt to evolving channel conditions and deployment scenarios. In this demo, we showcase an \ac{ML}-based solution for dynamically generating new phase masks during runtime. Our approach leverages \iac{ML} model deployed on the microcontroller for approximating 
the output of a phase mask generation algorithm, responding to new inputs while remaining smaller than a codebook. 
 
\end{abstract}

\acresetall

\iftrue
\fancypagestyle{firstpage}
{
    \fancyhead[L]{This work has been submitted to the IEEE for possible
      publication.\\
      Copyright may be transferred without notice, after which this version may no longer be accessible.}
    \fancyhead[R]{}
    \pagenumbering{gobble}
}

\thispagestyle{firstpage}

\fi

\input{sections/01_introduction.tex}
\input{sections/02_demo_design.tex}

\input{sections/03_phase_mask_generation.tex}

\input{sections/04_numerical_validation}

\section*{Acknowledgements}

The research leading to this paper received support from the Commonwealth 
Cyber Initiative. 
For more information about CCI, visit: 
\url{www.cyberinitiative.org}.
This material is also based upon work supported by the National Science 
Foundation under Grants Nos. 2318798 and 2326599.  
The authors also thank InterDigital for providing them with their mmWave equipment.

\bibliographystyle{IEEEtran}
\bibliography{IEEEabrv,bibliography}


\end{document}

%% file: sections/01_introduction.tex
\section{Introduction}\label{sec:intro}

\ac{mmWave} communication systems are prone to increased path loss and susceptible to obstructions in the environment, which can reduce the \ac{RSS} and potentially lead to link degradation or failure~\cite{rappaport2012broadband}. To address these challenges, 
\acp{RIS} can be deployed in the environment to improve the coverage and reliability of communication
by redirecting, reflecting, or reshaping incident signals when the direct link is weak or blocked~\cite{nemati2020ris}.
\acp{RIS} are passive metasurfaces that operate as electronically steerable reflectors, reducing energy consumption and electromagnetic footprint compared to active relays or repeaters. 

The reflection pattern of a \ac{RIS} is determined by the phase shifts of its radiating elements.
These phase shifts are configured through \textit{phase masks}, which are applied to all elements by low-power embedded microcontrollers~\cite{Shekhawat2025}. 
However, obtaining an optimal phase mask requires computationally intensive generation algorithms and an extensive amount of high-accuracy channel geometry data, especially as the size of the \ac{RIS} increases~\cite{zhou2023survey}.
As a result, prototype and commercial \ac{RIS} models, such as~\cite{Shekhawat2025} and~\cite{tmytek_xrifle_dynamic_ris_28g}, respectively, often rely on lookup tables (known as codebooks) that store predefined phase masks generated offline for representative reflection configurations. These phase masks can then be iteratively applied during beam training, or retrieved directly based on user-defined steering angles or locations~\cite{osman2025realtime}.

\begin{figure}[t]
    \centering
    \includegraphics[width=0.99\columnwidth]{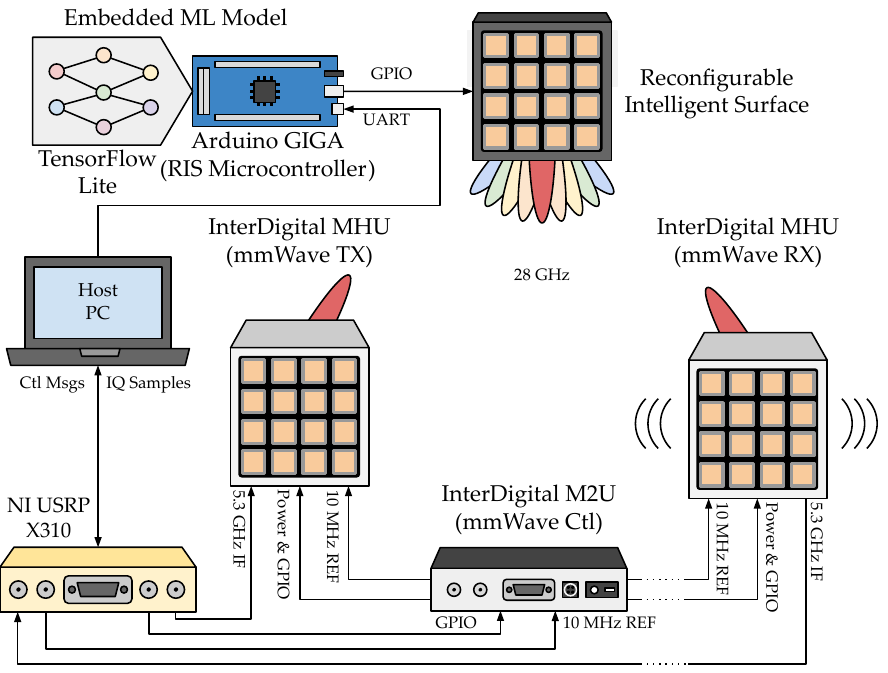}
    \caption{Demo setup with an \ac{RIS} and two mmWave radios. A movable RX reports \ac{RSS} to the Host PC, which performs adaptive beam management on the \ac{RIS}, and our embedded \ac{ML} model dynamically generates phase masks to steer reflected signals toward the RX.}
    \label{fig:arch}
    \vspace{-2em}
\end{figure}

The key limitation of this approach is that pregenerating a codebook to cover all foreseeable channel conditions (e.g., the presence of specific obstacles), deployment scenarios (i.e., far-field or near-field), and use cases (e.g., communications, ranging) may be intractable and far exceed the storage capacity of the embedded microcontrollers, hindering the deployment of \acp{RIS} in practical settings~\cite{abdallah2024multi}.
For instance, an Arduino Mega microcontroller has \unit[256]{KB} of flash storage, and it could (ideally) store nearly 2000 phase masks, as each phase mask for a 1024-element 1-bit \ac{RIS} occupies at least \unit[128]{B}.
While this may appear large, 2000 masks provide only up to 13 discrete positions per dimension when 
uniformly distributed across three spatial dimensions, before considering different channel conditions, obstacles, or use cases. 
To address this challenge, we propose to replace the codebook with \iac{ML} model, implemented as a lightweight \ac{NN} embedded on the \ac{RIS} microcontroller, for approximating the output of the phase-mask generation algorithm and enabling a flexible generation of phase masks during runtime.

In this demo, we showcase the first experimental solution for dynamically generating phase masks in support of adaptive \ac{RIS}-aided \ac{mmWave} communication systems.
Our solution comprises \iac{ML} model trained on an extensive set of pregenerated near-field phase masks for a 1-bit \ac{RIS}, embedded into the microcontroller using TensorFlow Lite, a framework for compressing \ac{ML} models and running them in inexpensive microcontrollers~\cite{warden2019tinyml}. 
This demonstration will show how dynamic phase-mask generation enables an adaptive \ac{RIS}-aided communication system that is not constrained by fixed codebooks and can adapt to different deployment scenarios. By generating new phase masks at runtime, the system performs adaptive beam management on the \ac{RIS}, tracking radio locations and steering the reflected signal toward new RX positions to mitigate link degradation due to mobility.

%% file: sections/02_demo_design.tex
\section{Demo Design and Functionality}\label{sec:arch}

Fig.~\ref{fig:arch} shows our demo setup, where TX and RX \ac{mmWave} radios communicate over a non-line-of-sight link aided by \iac{RIS}. 
Our demo leverages \iac{RIS} developed by Arizona State University~\cite{Shekhawat2025}, \ac{mmWave} equipment provided by InterDigital, and resources from the CCI xG Testbed~\cite{da2025cci}, which include: 2x \ac{MHU} mmWave front-ends, 1x  M2U mmWave controller, 1x USRP X310 \ac{SDR}, and 1x Host PC. 
The \acp{MHU} represent the TX and RX radios, controlled by the accompanying \ac{M2U} unit, which receives control signals from an \ac{SDR}, while the \ac{RIS} is controlled by the Host PC.
Each \ac{MHU} contains an $8 \times 8$ phased array antenna configured to boresight, and up/down-converts signals with bandwidth up to 200 MHz from/to \ac{IF} 5.3 GHz to/from the 28 GHz band, with a transmit power of \unit[37]{dBm}. This \ac{IF} allows the \acp{MHU} to interface with a commercial FR1 \ac{SDR} through a coaxial cable. We implemented the mmWave control and adaptive beam management on top of STAMINA, an out-of-tree GNU Radio module developed in our previous works~\cite{santos2023stamina, baron2023adaptive, baron2024eavesdropper}, which we extended with new blocks for interfacing with the \ac{RIS} via UART. 

The \ac{RIS} contains 1024 1-bit elements arranged in a $32 \times 32$ grid, allowing each radiating element to apply a phase shift of $0^\circ$ or $180^\circ$. 
We control the phase of each radiating element using an Arduino GIGA microcontroller running custom firmware. The microcontroller loads a phase mask into memory, converts it into a bitstream format, and shifts the corresponding bits into shift registers on the \ac{RIS} to configure all elements at the same time. To generate the phase masks, we have embedded an ML model trained using PyTorch and converted to TensorFlow Lite for deployment on the Arduino. 
The system performs beam tracking by updating the \ac{RIS} phase masks based on \ac{RSS} feedback from the RX, steering the reflected signal toward the RX as its position changes. Each position change event is communicated to the \ac{RIS} through a serial connection via UART and triggers the phase-mask generation performed by our proposed method.

%% file: sections/03_phase_mask_generation.tex
\section{\ac{ML}-based Generation of Phase Masks}\label{sec:masks}

\begin{figure}[t]
    \centering
    \includegraphics[width=0.95\columnwidth]{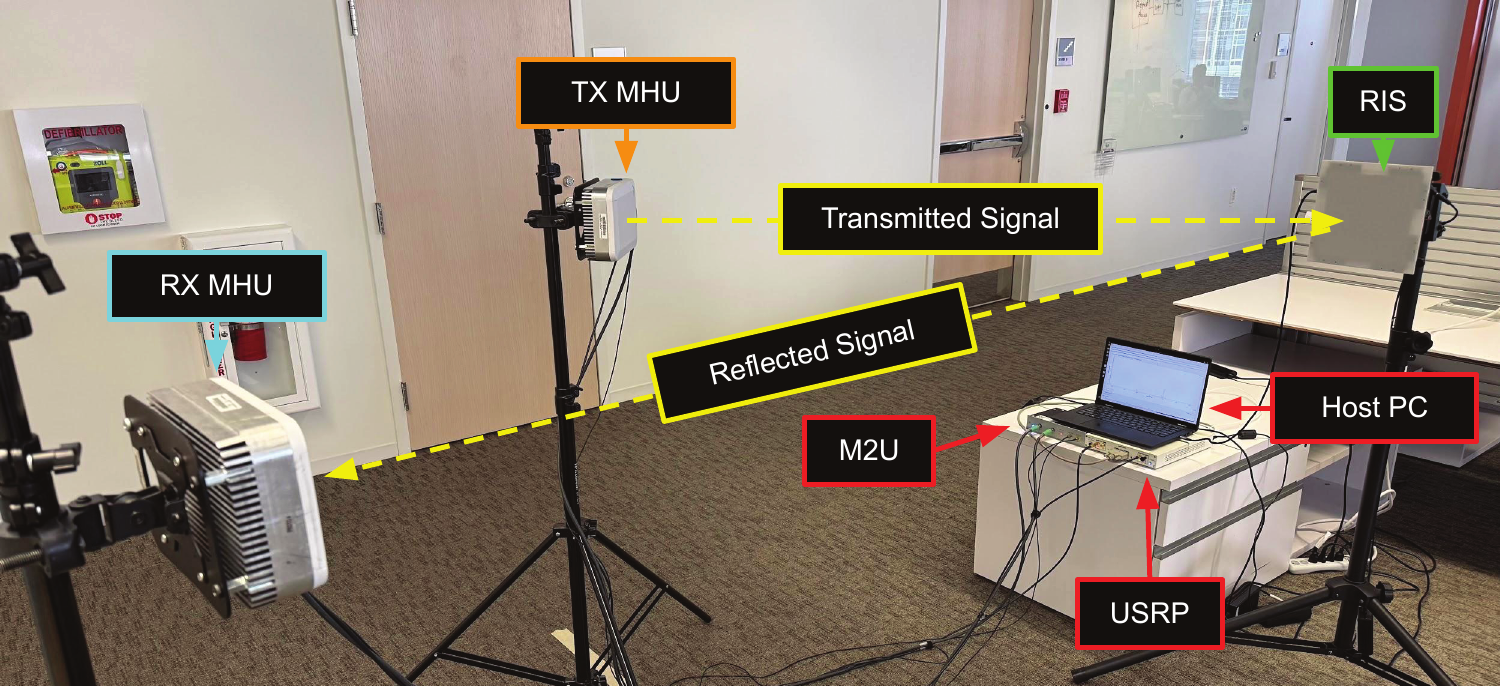}
    \caption{Experimental setup we used to develop and evaluate our \ac{ML}-based solution for dynamically generating phase masks during runtime to steer the reflected signals of a \ac{RIS}.}
    \label{fig:photo}
\end{figure}

The traditional approach of pregenerating phase masks and storing them in a codebook limits the operation of an \ac{RIS} to known channel conditions and deployment scenarios~\cite{abdallah2024multi}. 
When the \ac{RIS} encounters conditions that require a mask not included in the current codebook, the new mask must be generated externally using a generation algorithm, added to the microcontroller firmware, and deployed by taking the \ac{RIS} offline and re-flashing its microcontroller. These steps add significant overhead when the \ac{RIS} needs to respond to new operating conditions.
To address these challenges, we embedded an \ac{ML} model on the \ac{RIS} microcontroller to generate new phase masks at runtime. 
This approach allows the \ac{RIS} to adapt to a broader range of operating conditions than would be possible with a fixed codebook, without taking the \ac{RIS} offline or updating its firmware.

Our \ac{NN} architecture consists of a four-layer, feedforward, fully connected \ac{MLP}. The input layer contains 128 neurons that encode the coordinates of the TX and RX radios as binary activation values. The two hidden layers contain 256 neurons each, and use ReLU activation. The output layer contains one neuron per element of the RIS. The input to the network consists of four 32-bit floating-point values representing the $(x,y)$ coordinates of the TX and RX radios in millimeters, relative to the \ac{RIS}. We convert these values to binary and use their 128-bit concatenation as the input vector.
As we are utilizing a 1-bit \ac{RIS} in our demo, we binarize the output layer by mapping positive activations to 1 and negative activations to 0.
The same approach can be adapted to \acp{RIS} with higher phase resolution by quantizing the output into $2^b$ phase states for a $b$-bit \ac{RIS}, 
or by directly using the output for continuously tunable \acp{RIS}.
We then use the resulting binary output as an approximate phase mask to control the \ac{RIS} and steer the direction of its reflected signals.

We leveraged an analytical algorithm developed at Arizona State University~\cite{Shekhawat2025} to generate a dataset of phase masks corresponding to different 
positions of the radios.
The algorithm takes as input the  $(x,y,z)$ coordinates of the RX and TX, and calculates the required phase shift that each element needs to impose on the incident signal from the TX, to allow the overall beam to be steered towards the RX position.
The phase shifts are quantized to $0^\circ$ and $180^\circ$ and translated to binary values of 0 and 1, respectively, for use by the 1-bit RIS. 
The final output is a grid of $32 \times 32 = 1024$ bits, which are distributed to the radiating elements through a circuit of shift registers discussed in~\cite{Shekhawat2025}. 
We generated multiple phase masks corresponding to different positions of the TX and RX radios, and combined them into a labeled dataset based of their locations relative to the \ac{RIS} (in millimeters).
We used this dataset for training our \ac{ML}-based solution for dynamically generating phase masks, and 
to serve as the ground-truth phase masks in our evaluation.

%% file: sections/04_numerical_validation.tex
\begin{figure}[t]
    \centering
    \includegraphics[width=0.99\columnwidth]{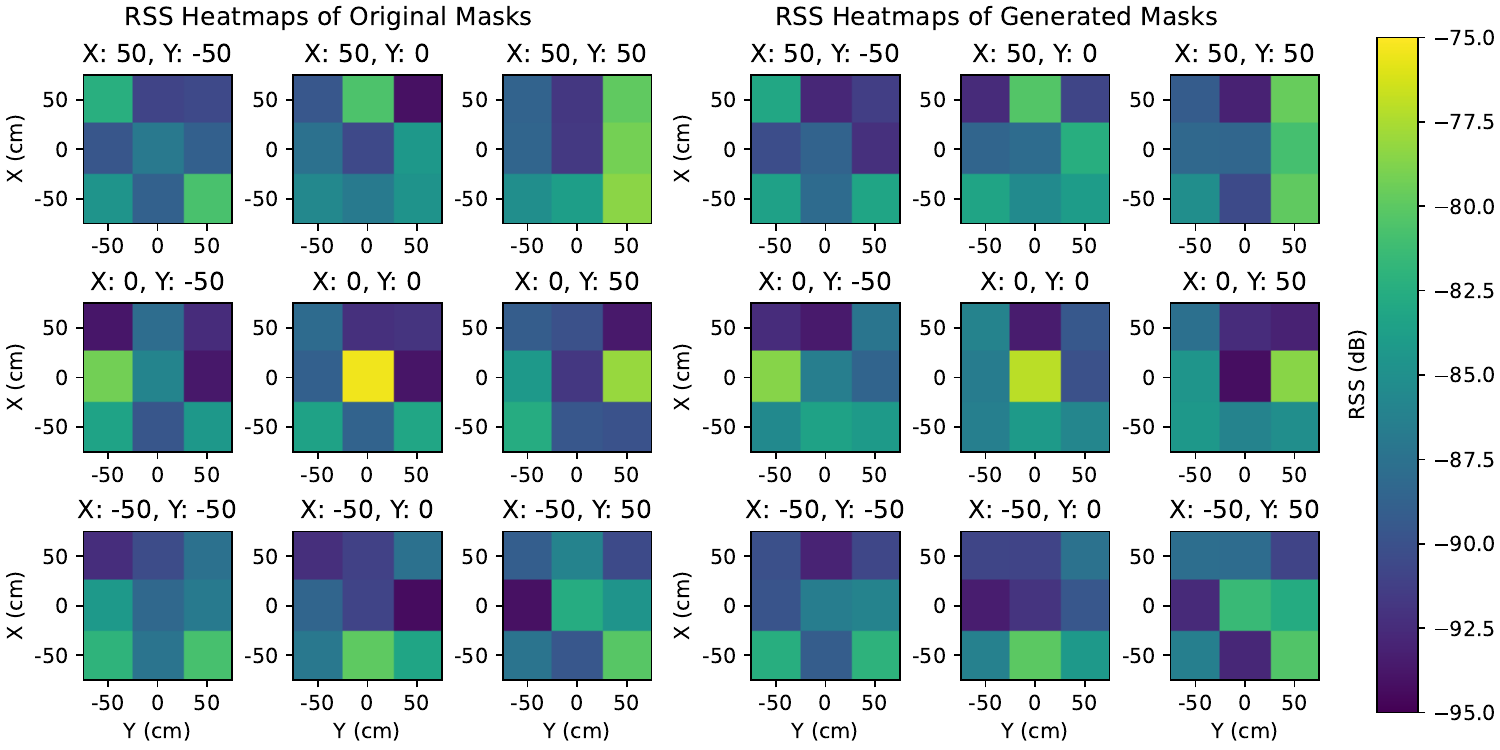}
    \caption{Heatmaps showing the average \ac{RSS} collected at each RX location measured over \unit[10]{s}, comparing the original phase masks (left) and our \ac{ML}-generated phase masks (right).
    The heatmaps are arranged according to each phase mask's target steering location in the $(x,y)$ evaluation grid.}
    \label{fig:measurements}
\end{figure}

\section{Numerical Validation}

To validate our solution for dynamically generating phase masks, we trained our \ac{ML} model on a dataset representing a non-line-of-sight scenario where the communication between TX and RX is aided by the \ac{RIS}.
To do so, we created a dataset containing 1681 near-field phase masks, corresponding to a $41 \times 41$ grid of different RX positions on the $(x,y)$ dimensions (spaced \unit[5]{cm} apart),  located at a fixed $z = \unit[2]{m}$, where $z$ denotes the distance from the \ac{RIS} plane. 
The TX was fixed at \unit[1]{m} from the \ac{RIS} along $z$ and \unit[1]{m} to its right along $x$.
Then, we evaluated our solution on a coarser experimental setup, where the RX was moved to different locations on a $3\times3$ evaluation grid on the $(x,y)$ plane centered on boresight, with \unit[50]{cm} spacing, as shown in Fig.~\ref{fig:photo}.
For this initial validation, our \ac{NN} architecture supports two spatial degrees of freedom, corresponding to movement along the $x$ and $y$ dimensions at a fixed distance $z$. In future work, we plan to extend our architecture to also support a variable $z$, enabling dynamic phase-mask generation across different distances.

At each RX location, we compared the \ac{RSS} obtained from the nine ground-truth phase masks and their nine \ac{ML}-generated counterparts, each targeting one RX location in the $(x,y)$ evaluation grid, as shown in Fig.~\ref{fig:measurements}.
As expected, our \ac{ML}-based solution generates phase masks that closely mimic the behavior of the ground-truth masks, steering reflected signals toward desired locations,  
as well as reproducing unexpected effects from the training dataset, such as the secondary reflections shown in the top right location $(x,y) =( 50,50)$. 
In addition, the reflected signals displayed similar performance at the target steering locations, with the average \ac{RSS} obtained using the generated phase masks being only \unit[0.26]{dB} lower on average than the ground-truth phase masks.
These results demonstrate that our \ac{ML}-generated masks perform very similarly to the ground-truth masks and validate our \ac{ML}-based solution for dynamically generated phase masks.

During the demonstration, attendees will be able to move the RX to different locations and observe, through a dashboard, an adaptive \ac{RIS}-aided communication link enabled by our approach for dynamically generating phase masks.
The movable RX will report \ac{RSS} measurements to a Host PC, which performs adaptive beam management on the \ac{RIS} in GNU Radio and sends updated RX coordinates to the microcontroller. The embedded \ac{ML} model will then generate phase masks to steer the reflected signal toward the updated RX position. 
The dashboard will display the \ac{RSS}, the generated mask, and its array factor, illustrating how the system can mitigate performance degradation due to mobility.